\documentclass[aps,pra,showpacs,twocolumn]{revtex4}
\usepackage{graphicx}
\begin{document}

\title[Prospects for observing DCE and Anti-DCE in circuit QED]{Prospects
for observing dynamical and anti- dynamical Casimir effects in circuit QED
due to fast modulation of qubit parameters}
\author{D S Veloso$^1$ and A V Dodonov$^{1,2}$}
\affiliation{$^1$ Institute of Physics,
University of Brasilia, 70910-900, Brasilia, Federal District, Brazil}

\affiliation{$^2$ International Center for Condensed Matter Physics,
University of Brasilia, 70910-900, Brasilia, Federal District, Brazil}

\begin{abstract}
We consider the nonstationary circuit QED architecture, where a single
artificial two-level atom interacts with a cavity field mode under external
modulation of one or more system parameters. Two different approaches are
employed to study the effects of Markovian dissipation on modulation-induced
transitions between the atom--field dressed states: the standard master
equation of Quantum Optics and the recently formulated dressed-picture
master equation. We estimate the associated transition rates and show that
photon generation from vacuum (\textquotedblleft dynamical Casimir
effect\textquotedblright , DCE) and coherent photon annihilation from
nonvacuum states (\textquotedblleft Anti-DCE\textquotedblright ) are
possible with the current state-of-the-art parameters.
\end{abstract}

\pacs{42.50.Pq, 42.50.Ct, 42.50.Hz, 32.80-t, 03.65.Yz}
\maketitle

\section{Introduction}

The subject of photon generation in nonstationary circuit QED, where
mesoscopic \textquotedblleft artificial atoms\textquotedblright\
(constructed with Josephson junctions \cite{circ1,circ2}) interact with the
Electromagnetic field confined in superconducting stripline resonators under
nonstationary conditions \cite{nori}, has been studied theoretically for
nearly ten years. Traditionally one considered the externally prescribed
modulation of either the cavity frequency \cite%
{pla,fedotov,roberto,2level,Zei,2,sinaia1,sinaia2,cacheffo}, the atom--field
coupling strength \cite{liberato7,liberato9,como,como1,solano2} or the
atomic transition frequency \cite{LZ1,LZ2,arxiv,jpcs,werlang,vedral}, while
recently more sophisticated schemes involving multi-level atoms, several
qubits or coupled cavities were examined \cite%
{3level,back,2atom,sinaia1,camop,solano}. Such a unique possibility of \emph{%
in situ} control of the fundamental parameters of the Hamiltonian became
possible thanks to the highly controllable solid state environment in which
the system properties can be manipulated by external electric and magnetic
fields \cite{cir1,cir2,nori-n,meta,joha}. Recently it was shown that
modulation of \emph{any} system parameter gives rise to similar effects of
roughly the same order of magnitude, and simultaneous modulation of
different parameters with the same frequency can increase or decrease the
associated transition rate depending on the relative phases of modulations
\cite{I1,I2}. Moreover, \textquotedblleft multi-tone
modulations\textquotedblright\ comprising two or more harmonic functions of
time can lead to new types of effective interactions \cite{jpcs,I1,I2}.

However, actual circuit QED architectures suffer from unavoidable
dissipation effects, in particular, the damping of the cavity field, the
atomic relaxation and the pure atomic dephasing \cite{blais}. Although some
works on photon generation in nonstationary circuit QED analyzed the effects
of losses \cite{liberato7,liberato9,roberto,back,cache}, the majority of
studies investigated only the unitary dynamics for different regimes of
parameters and modulation shapes. Now that the general analytical
description in absence of losses has been formulated in a closed form (under
a series of approximations) \cite{I1,I2}, the question of utmost practical
interest is how dissipation affects the modulation-induced phenomena and
whether they can be implemented with present or near-future technology.

So the goal of this work is to unveil which nonstationary phenomena can be
implemented experimentally with the current technology \cite{art1,art2,art3}
and comprehend how different dissipation channels affect the time evolution.
We concentrate on the single-qubit, single-mode circuit QED setup and take
into account the three aforementioned dissipation mechanisms using the
Markovian and zero-temperature approximations. The master equation
appropriate to this case, which takes into account the qubit--resonator
coupling, was deduced microscopically in \cite{blais}. In this so called
\textquotedblleft dressed-picture master equation\textquotedblright\ (DPME)
the dissipation rates depend on the spectrum of noise evaluated at the
frequency matching the transition between the atom--field eigenstates
(dressed states), so a prior knowledge of the bath structure is required.
Besides, the analytical solution is complicated in majority of cases, so we
carry out exact numerical simulations by assuming that the noise has a flat
spectrum at positive frequencies and is zero for the negative ones. The
details of the underlying mathematical formalism can be found in section \ref%
{s2}.

We consider several regimes of photon generation and annihilation due to
modulation of the atomic transition frequency and the atom--field coupling
strength. First we study the generation of two excitations from vacuum in
the resonant regime (section \ref{s4c}) and the creation of one photon and
one atomic excitation in the dispersive regime (\emph{Anti-Jaynes-Cummings}
behavior, section \ref{s4a}) for single-tone modulations. Then we analyze
the recently discovered \emph{Anti-dynamical Casimir effect} (Anti-DCE,
section \ref{s4b}), whereby two excitations are coherently annihilated due
to the modulation of qubit parameters in the dispersive regime \cite{I2}. We
show that the above phenomena can be enhanced by using two-tone modulations
(section \ref{s5}), for which a total of four excitations can be generated
from vacuum or annihilated from a known initial state. Quantum states with
even more excitations can be generated from vacuum in the dispersive regime
via the \emph{dynamical Casimir effect} (DCE, section \ref{s3}), when the
system parameters are modulated with frequency roughly equal to twice the
cavity unperturbed frequency.

It is shown that all the above phenomena can be implemented with current
technology in dissipative circuit QED, although DCE and Anti-DCE require
state-of-the-art dissipation rates and precise tuning of the modulation
frequency. Moreover, we demonstrate numerically that for time intervals of
interest the predictions of DPME are almost indistinguishable from the ones
obtained for the \textquotedblleft standard master
equation\textquotedblright\ (SME) of Quantum Optics \cite{vogel,schleich},
which is much simpler to handle and whose approximate solution is deduced
here in the long-time limit.

\section{Mathematical formalism}

\label{s2}

The atom--field interaction is described by the Rabi Hamiltonian \cite%
{rabi,jpcs,werlang} and for generality we take into account the parametric
amplification term due to eventual time-modulation of the cavity parameters
\cite{law94}. The total Hamiltonian reads (we set $\hbar =1$)%
\begin{equation}
\hat{H}=\omega \hat{n}+\Omega |e\rangle \langle e|+g(\hat{a}+\hat{a}%
^{\dagger })(\hat{\sigma}_{+}+\hat{\sigma}_{-})+i\chi (\hat{a}^{\dagger 2}-%
\hat{a}^{2})\,,  \label{H1}
\end{equation}%
where $\hat{a}$ and $\hat{a}^{\dagger }$ are cavity annihilation and
creation operators and $\hat{n}=\hat{a}^{\dagger }\hat{a}$ is the photon
number operator; $\hat{\sigma}_{+}=|e\rangle \langle g|$ and $\hat{\sigma}%
_{-}=|g\rangle \langle e|$ are the atomic ladder operators, where $|g\rangle
$ ($|e\rangle $) denotes the atomic ground (excited) state. $\omega $ is the
cavity frequency, $\Omega $ is the atomic transition frequency, $g$ is the
atom--field coupling strength and $\chi $ is the squeezing coefficient
related to the parametric amplification process. The time-independent part
of $\chi $ may appear due to the terms proportional to the square of the
vector potential \cite{schleich}, while the time-dependent part is due to
the time-modulation of the cavity frequency and in the simplest case of DCE
reads $\chi =(4\omega )^{-1}d\omega /dt$.

In this work we follow the convention of papers \cite{I1,I2} and suppose
that all the system parameters can be perturbed simultaneously by external
modulations of the form $X=X_{0}+\varepsilon _{X}\sum_{j}w_{X}^{(j)}\sin
(\eta ^{(j)}t+\phi _{X}^{(j)})$. Here $X=\{\omega ,\Omega ,g,\chi \}$, $%
X_{0} $ and $\varepsilon _{X}\geq 0$ are the corresponding bare values and
modulation depths, and the sum runs over all the fast modulation frequencies
$\eta ^{(j)}>\omega _{0}$. We consider small modulation depths given by
inequalities $\varepsilon _{\omega },\varepsilon _{\Omega },\varepsilon _{g}%
\sqrt{n_{\max }},\varepsilon _{\chi }n_{\max }\ll \omega _{0}$, where $%
n_{\max }$ is the maximum number of excitations. The parameters $%
w_{X}^{(j)}\geq 0$ and $\varphi _{X}^{(j)}$ are the relative weights and
phase constants corresponding to the modulation of parameter $X$ at
frequency $\eta ^{(j)}$. To shorten the notation we introduce the complex
modulation depth $\varepsilon _{X}^{(j)}\equiv \varepsilon
_{X}w_{X}^{(j)}\exp (i\phi _{X}^{(j)})$ that incorporates both the weight
and the phase for the modulation frequency $\eta ^{(j)}$. In our numerical
examples the phases $\phi _{X}^{(j)}$ will always be set to zero and $%
w_{X}^{(j)}=1$ for single-tone modulations with frequency $\eta ^{(j)}$.
Whenever more than one parameter are modulated simultaneously we shall call
the process \textquotedblleft multi-modulation\textquotedblright.

We assume weak atom--field coupling, $g_{0}\sqrt{n_{\max }}\ll \omega _{0}$,
and $\chi _{0}=0$, so one can describe the dynamics in the basis of
eigenstates of the bare Jaynes-Cummings Hamiltonian (JCH) $\hat{H}%
_{JC}=\omega _{0}\hat{n}+\Omega _{0}|e\rangle \langle e|+g_{0}(\hat{a}\hat{%
\sigma}_{+}+\hat{a}^{\dagger }\hat{\sigma}_{-})$ \cite{schleich,vogel}.
These states, known as \emph{dressed states}, read%
\begin{equation}
|\varphi _{n,\mathcal{S}}\rangle ={\rm{s}}_{n,\mathcal{S}}|g,n\rangle +{\rm{%
c}}_{n,\mathcal{S}}|e,n-1\rangle ~,~n\geq 0~,~\mathcal{S}=\pm ~,~
\end{equation}%
where $n$ is the total number of system excitations and the index $\mathcal{S%
}$ labels different eigenstates with same $n$. We define formally $|\varphi
_{0,+}\rangle \equiv 0$ and introduce the notation ${\rm{s}}_{m,-}={\rm{c}}%
_{m,+}=\cos \theta _{m}$, ${\rm{s}}_{m,+}=-{\rm{c}}_{m,-}=\sin \theta _{m}$,
where $\theta _{0}=0$, $\theta _{n>0}=\arctan [(\Delta _{-}+\beta
_{n})/(2g_{0}\sqrt{n})]$ and $\beta _{n}=\sqrt{\Delta _{-}^{2}+4g_{0}^{2}n}$%
. The eigenenergies are $\lambda _{0,-}=0$ and $\lambda _{n,\mathcal{S}%
}=\omega _{0}n-\Delta _{-}/2+\mathcal{S}\beta _{n}/2$, where $\Delta
_{-}=\omega _{0}-\Omega _{0}$ is the bare detuning (we assume $|\Delta
_{-}|\ll \omega _{0}$).

The approximate unitary dynamics for the Hamiltonian (\ref{H1}) was solved
in \cite{I1,I2} for arbitrary small-depth modulations. In the interaction
picture defined by the unitary transformation $\hat{U}_{t}=\exp (-it\hat{H}%
_{JC})$ the effective Hamiltonian reads%
\begin{eqnarray}
\tilde{H}&=&\sum_{m,\mathcal{S},\mathcal{T}}\sum_{j}\Xi _{m,\mathcal{T},%
\mathcal{S}}^{(j)}e^{-it(\lambda _{m+2,\mathcal{S}}-\lambda _{m,\mathcal{T}%
}-\eta ^{(j)})} \nonumber \\
&&\times |\varphi _{m,\mathcal{T}}\rangle \langle \varphi _{m+2,%
\mathcal{S}}|+h.c.~,  \label{tild}
\end{eqnarray}%
where $m\geq 0$, $\mathcal{S},\mathcal{T}=\pm $ and the time-independent
coefficients $\Xi _{m,\mathcal{T},\mathcal{S}}^{(j)}$ will be given
throughout the paper (the interaction-picture density operator is $\tilde{%
\rho}=\hat{U}_{t}^{\dagger }\hat{\rho}\hat{U}_{t}$). By choosing $\eta
_{M}=\lambda _{M+2,\mathcal{S}}-\lambda _{M,\mathcal{T}}$ one resonantly
couples the dressed states $|\varphi _{M,\mathcal{T}}\rangle \leftrightarrow
|\varphi _{M+2,\mathcal{S}}\rangle $. If in addition $|\lambda _{m+2,%
\mathcal{S}}-\lambda _{m,\mathcal{T}}-\eta _{M}|\gg |\Xi _{m,\mathcal{T},%
\mathcal{S}}^{(j)}|$ for other values of $\{m,\mathcal{S},\mathcal{T}\}$,
all other terms can be neglected under the Rotating Wave Approximation (RWA)
\cite{I1,I2} and one obtains a simplified time-independent Hamiltonian.
Actually, the eigenenergies $\lambda _{n,\mathcal{S}}$ must be corrected by
the so called \textquotedblleft intrinsic frequency
shifts\textquotedblright\ of the order ${\rm{O}}(g_{0}^{2}/\omega
_{0},\varepsilon _{X}^{2}/\omega _{0})$ with $X=\{\omega ,\Omega ,g,\chi \}$%
. These corrections do not alter significantly the mathematical formalism,
so we neglect them in the analytical derivations below; however, they will
be incorporated into our numerical simulations to achieve exact resonances.

In the presence of dissipation the dynamics must be described by the master
equation for the density operator%
\begin{equation}
d\hat{\rho}/dt=-i[\hat{H},\hat{\rho}]+\mathcal{\hat{L}}\hat{\rho},
\end{equation}%
where $\mathcal{\hat{L}}$ is the Liouvillian superoperator. We shall use two
different Markovian kernels to estimate the effects of dissipation. The
quantum optical \textquotedblleft standard master
equation\textquotedblright\ (SME) at zero temperature is \cite{vogel}%
\begin{equation}
\mathcal{\hat{L}}_{SME}\bullet =\kappa \mathcal{D}[\hat{a}]\bullet +\gamma
\mathcal{D}[\hat{\sigma}_{-}]\bullet +\frac{\gamma _{\phi }}{2}\mathcal{D}[%
\hat{\sigma}_{z}]\bullet \,,
\end{equation}%
where $\mathcal{D}[\hat{O}]\hat{\rho}\equiv \frac{1}{2}(2\hat{O}\hat{\rho}%
\hat{O}^{\dagger }-\hat{O}^{\dagger }\hat{O}\hat{\rho}-\hat{\rho}\hat{O}%
^{\dagger }\hat{O})$ is the Lindbladian superoperator and $\hat{\sigma}%
_{z}=|e\rangle \langle e|-|g\rangle \langle g|$. The constant parameters $%
\kappa $, $\gamma $ and $\gamma _{\phi }$ denote the cavity damping, qubit
relaxation and the qubit pure dephasing rates, respectively.

The second approach is the \textquotedblleft dressed-picture master
equation\textquotedblright\ (DPME) at zero temperature developed in \cite%
{blais}%
\begin{eqnarray}
\mathcal{L}_{dr}\bullet &=&\mathcal{D}\left[ \sum_{l}\Phi ^{l}|l\rangle
\langle l|\right] \bullet +\sum_{l,k\neq l}\Gamma _{\phi }^{lk}\mathcal{D}%
\left[ |l\rangle \langle k|\right] \bullet  \nonumber \\
&&+\sum_{l,k>l}\left( \Gamma _{\kappa }^{lk}+\Gamma _{\gamma }^{lk}\right)
\mathcal{D}\left[ |l\rangle \langle k|\right] \bullet \,,  \label{bla}
\end{eqnarray}%
where we use the shorthand notation $|l\rangle $ to denote the JCH\ dressed
states $|\varphi _{n,\mathcal{S}}\rangle $, $l$ increasing with the energy $%
\lambda _{n,\mathcal{S}}$. The parameters of equation (\ref{bla}) are
defined as $\Phi ^{l}=[\gamma _{\phi }(0)/2]^{1/2}\sigma _{z}^{ll}$,$%
~~\Gamma _{\phi }^{lk}=\gamma _{\phi }(\Delta _{kl})|\sigma _{z}^{lk}|^{2}/2$%
, $\Gamma _{\kappa }^{lk}=\kappa (\Delta _{kl})|a^{lk}|^{2}$ and$~\Gamma
_{\gamma }^{lk}=\gamma (\Delta _{kl})|\sigma _{x}^{lk}|^{2}$. Here $\kappa
(\nu )$, $\gamma (\nu )$ and $\gamma _{\phi }(\nu )$ are the dissipation
rates corresponding to the resonator and qubit dampings and dephasing noise
spectral densities at frequency $\nu $; we also defined $\Delta
_{kl}=\lambda _{k}-\lambda _{l}$, $\sigma _{z}^{lk}=\langle l|\hat{\sigma}%
_{z}|k\rangle $, $a^{lk}=\langle l|(\hat{a}+\hat{a}^{\dagger })|k\rangle $
and $\sigma _{x}^{lk}=\langle l|(\hat{\sigma}_{+}+\hat{\sigma}_{-})|k\rangle
$. We do not consider a specific model for the reservoirs and make the
simplest assumption that for $\nu \geq 0$ the dissipation rates are constant
and equal to the corresponding rates of SME, while for $\nu <0$ they are
zero.

We shall solve both master equations exactly via numerical integration of
the differential equations for the density matrix elements. We assume the
usual value $\omega _{0}/2\pi =8\,$GHz for the cavity frequency, high but
feasible value $g_{0}=5\times 10^{-2}\omega _{0}$ for the qubit--field
coupling strength (within the weak coupling regime), moderate range of
detuning $|\Delta _{-}|\leq 8g_{0}$ and the state-of-the-art dissipation
rates $\kappa \sim \gamma \sim \gamma _{\phi }\sim 5\times 10^{-5}g_{0}$
\cite{art1,art2,art3}. We presume the tentative value $|\Upsilon |\sim
10^{-2}$ for the collective relative modulation depth [$\Upsilon \sim
\sum_{X}\varepsilon _{X}^{(j)}/X_{0}$, see equations (\ref{hob1}), (\ref%
{hob2}), (\ref{hob3}), (\ref{hob4}) below], small enough to fulfill the
range of validity of the Hamiltonian (\ref{tild}) yet sufficient to
implement the phenomena of interest.

\subsection{Simplifications for the standard master equation}

The interaction-picture density operator obeys the master equation%
\begin{equation}
\frac{d\tilde{\rho}}{dt}=-i[\tilde{H},\tilde{\rho}]+\kappa \mathcal{D}[%
\tilde{a}]\tilde{\rho}+\gamma \mathcal{D}[\tilde{\sigma}_{-}]\tilde{\rho}+%
\frac{\gamma _{\phi }}{2}\mathcal{D}[\tilde{\sigma}_{z}]\tilde{\rho},
\label{kennedy}
\end{equation}%
where $\tilde{O}\equiv \hat{U}_{t}^{\dagger }\hat{O}\hat{U}_{t}$. The
natural basis to expand $\tilde{\rho}$ consists of the JCH dressed states%
\begin{equation}
\tilde{\rho}(t)=\sum_{n,m=0}^{\infty }\sum_{\mathcal{S},\mathcal{T}=\pm }%
\tilde{\rho}_{n,m}^{\mathcal{S},\mathcal{T}}(t)|\varphi _{n,\mathcal{S}%
}\rangle \langle \varphi _{m,\mathcal{T}}|~.  \label{slash}
\end{equation}

By substituting equation (\ref{slash}) into (\ref{kennedy}) one can obtain
the differential equations for the density matrix elements $\tilde{\rho}%
_{n,m}^{\mathcal{S},\mathcal{T}}(t)$. For example, for the pure atomic
dephasing we get [using $\tilde{\rho}_{n,m}^{-,+}=(\tilde{\rho}%
_{m,n}^{+,-})^{\ast }$]%
\begin{eqnarray}
\frac{d\tilde{\rho}_{N,M}^{+,+}}{dt} &=&\frac{\gamma _{\phi }}{2}\left[
\left( \cos 2\theta _{M}\cos 2\theta _{N}-1\right) \tilde{\rho}%
_{N,M}^{+,+}\right.  \label{1} \\
&&-e^{-it\beta _{M}}\sin 2\theta _{M}\cos 2\theta _{N}\tilde{\rho}%
_{N,M}^{+,-}  \nonumber \\
&&-e^{it\beta _{N}}\sin 2\theta _{N}\cos 2\theta _{M}(\tilde{\rho}%
_{M,N}^{+,-})^{\ast }  \nonumber \\
&&\left. +e^{it\left( \beta _{N}-\beta _{M}\right) }\sin 2\theta _{M}\sin
2\theta _{N}\tilde{\rho}_{N,M}^{-,-}\right]  \nonumber
\end{eqnarray}%
\begin{eqnarray}
\frac{d\tilde{\rho}_{N,M}^{-,-}}{dt} &=&\frac{\gamma _{\phi }}{2}\left[
\left( \cos 2\theta _{M}\cos 2\theta _{N}-1\right) \tilde{\rho}%
_{N,M}^{-,-}\right.  \label{2} \\
&&+e^{-it\beta _{N}}\sin 2\theta _{N}\cos 2\theta _{M}\tilde{\rho}%
_{N,M}^{+,-}  \nonumber \\
&&+e^{it\beta _{M}}\sin 2\theta _{M}\cos 2\theta _{N}(\tilde{\rho}%
_{M,N}^{+,-})^{\ast }  \nonumber \\
&&\left. +e^{-it\left( \beta _{N}-\beta _{M}\right) }\sin 2\theta _{M}\sin
2\theta _{N}\tilde{\rho}_{N,M}^{+,+}\right]  \nonumber
\end{eqnarray}%
\begin{eqnarray}
\frac{d\tilde{\rho}_{N,M}^{+,-}}{dt} &=&-\frac{\gamma _{\phi }}{2}\left[
\left( \cos 2\theta _{M}\cos 2\theta _{N}+1\right) \tilde{\rho}%
_{N,M}^{+,-}\right.  \label{3} \\
&&+e^{it\beta _{M}}\sin 2\theta _{M}\cos 2\theta _{N}\tilde{\rho}_{N,M}^{+,+}
\nonumber \\
&&-e^{it\beta _{N}}\sin 2\theta _{N}\cos 2\theta _{M}\tilde{\rho}_{N,M}^{-,-}
\nonumber \\
&&\left. -e^{it\left( \beta _{N}+\beta _{M}\right) }\sin 2\theta _{M}\sin
2\theta _{N}(\tilde{\rho}_{M,N}^{+,-})^{\ast }\right] \,.  \nonumber
\end{eqnarray}

When one considers the pair of equations%
\begin{equation}
\frac{d}{dt}A(t)=-qe^{itw}B(t)~,~\frac{d}{dt}B(t)=-qe^{-itw}A(t)
\label{saul}
\end{equation}%
with constant parameters $q$ and $w$, the solution is%
\begin{eqnarray}
A &=&\frac{%
(w_{+}A_{0}-iqB_{0})e^{itw_{-}/2}-(w_{-}A_{0}-iqB_{0})e^{itw_{+}/2}}{r} \\
B &=&\frac{%
(w_{+}B_{0}+iqA_{0})e^{-itw_{-}/2}-(w_{-}B_{0}+iqA_{0})e^{-itw_{+}/2}}{r}~,
\nonumber
\end{eqnarray}%
where $r=\sqrt{w^{2}-4q^{2}}$ and $w_{\pm }\equiv (w\pm r)/2$. For $|q/w|\ll
1$ one has $A\simeq A_{0}$, $B\simeq B_{0}$, which is equivalent to $%
dA/dt\simeq 0$, $dB/dt\simeq 0$ in equation (\ref{saul}), provided one
neglects the \textquotedblleft frequency shifts\textquotedblright\ of the
order ${\rm{O}}(q^{2}/w)$. This observation constitutes the method of RWA \cite%
{I1,I2}, which allows one to neglect the rapidly oscillating terms in
differential equations for probability amplitudes. Hence, for $\beta _{N}\gg
\gamma _{\phi }$ one can neglect the second and the third lines in equations
(\ref{1}) -- (\ref{3}) and the fourth line in equation (\ref{3}). Moreover,
for nondiagonal elements one can also neglect the fourth line in equations (%
\ref{1}) -- (\ref{2}). Similar reasoning holds for the atomic and cavity
dampings.

Therefore, by choosing resonant modulation frequencies and performing RWA
one obtains time-independent coupled differential equations for the matrix
elements of $\tilde{\rho}$ that can be solved analytically or numerically.
The general solution can be quite cumbersome, so in this work we shall
pursue analytically only the asymptotic behavior, obtained by setting the
left-hand side of equations (\ref{1}) -- (\ref{3}) to zero.

\section{Two-excitations behavior}

\label{s4}

\subsection{Photon generation from vacuum in resonant regime}

\label{s4c}

First we study the photon creation from the initial zero-excitation state
(ZES) $|g,0\rangle $. In the resonant regime, $\Delta _{-}=0$, the resonant
modulation frequency was found two decades ago \cite{pla} and reads
\begin{equation}
\eta ^{(r)}=2\omega _{0}+\mathcal{R}g_{0}\sqrt{2}~,~\mathcal{R}=\pm \,.
\label{fre1}
\end{equation}%
The effective Hamiltonian is $\tilde{H}=\theta |\varphi _{0,-}\rangle
\langle \varphi _{2,\mathcal{R}}|+h.c.$, where%
\begin{equation}
\theta =ig_{0}\mathcal{R}\frac{\sqrt{2}}{4}\Upsilon ^{(r)}  \label{teta1}
\end{equation}%
\begin{equation}
\Upsilon ^{(r)}=\frac{\varepsilon _{\omega }^{(r)}}{2\omega _{0}}+\frac{%
\varepsilon _{\Omega }^{(r)}}{2\omega _{0}}-\frac{\varepsilon _{g}^{(r)}}{%
g_{0}}+\mathcal{R}i\sqrt{2}\frac{\varepsilon _{\chi }^{(r)}}{g_{0}}~.
\label{hob1}
\end{equation}%
Here $\Upsilon ^{(r)}$ denotes the dimensionless collective modulation depth
\cite{I2}. The analytical solution for this Hamiltonian in absence of losses
is straightforward [see equations (\ref{f1})--(\ref{f3}) below] and consists
of periodic oscillations between the states $|\varphi _{0,-}\rangle
\leftrightarrow |\varphi _{2,\mathcal{R}}\rangle $.

In the presence of dissipation described by SME, under the pure atomic
damping we obtain the nonzero asymptotic probabilities of dressed states ($%
\rho _{N}^{\mathcal{T}}\equiv \langle \varphi _{N,\mathcal{T}}|\hat{\rho}%
|\varphi _{N,\mathcal{T}}\rangle $)%
\begin{equation}
\rho _{0}^{\mathcal{-}}=\frac{\left\vert \theta \right\vert ^{2}+\left(
\gamma /4\right) ^{2}}{3\left\vert \theta \right\vert ^{2}+\left( \gamma
/4\right) ^{2}}\,,~\rho _{2}^{\mathcal{R}}=\frac{\left\vert \theta
\right\vert ^{2}}{3\left\vert \theta \right\vert ^{2}+\left( \gamma
/4\right) ^{2}}
\end{equation}%
and $\rho _{1}^{\pm }=\frac{1}{2}\rho _{2}^{\mathcal{R}}$. For the pure
atomic dephasing we obtain $\rho _{0}^{\mathcal{-}}=\rho _{2}^{\pm }=\frac{1%
}{3}$ and for the pure cavity relaxation%
\begin{eqnarray}
\rho _{0}^{\mathcal{-}} &=&\frac{\left\vert \theta \right\vert ^{2}+\left(
3\kappa /4\right) ^{2}}{5\left\vert \theta \right\vert ^{2}+\left( 3\kappa
/4\right) ^{2}}\,,~\rho _{2}^{\mathcal{R}}=\frac{\left\vert \theta
\right\vert ^{2}}{5\left\vert \theta \right\vert ^{2}+\left( 3\kappa
/4\right) ^{2}} \\
\rho _{1}^{\mathcal{R}} &=&\frac{1}{2}(\sqrt{2}+1)^{2}\rho _{2}^{\mathcal{R}%
}\,,~\rho _{1}^{-\mathcal{R}}=\frac{1}{2}(\sqrt{2}-1)^{2}\rho _{2}^{\mathcal{%
R}}\,.  \nonumber
\end{eqnarray}%
Hence dissipation affects dramatically the unitary dynamics, although
excitations can still be created whenever $\kappa ,\gamma \ll |\theta |$.

\begin{figure}[tbh]
\begin{center}
\includegraphics[width=.49\textwidth]{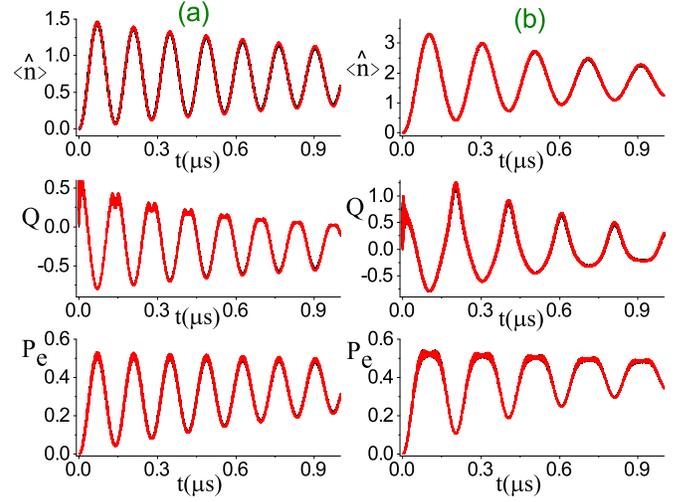} {}
\end{center}
\caption{Photon generation from ZES in the resonant regime.
\textbf{(a)} One-tone modulation (section \protect\ref{s4c}). \textbf{(b)}
Two-tone modulation (section \protect\ref{s5a}). Black (red) lines stand for
SME (DPME).}
\label{fig1}
\end{figure}

For our tentative parameters we obtain $|\theta |\sim 10^{-3}g_{0}$, so the
periodic generation of the state $|\varphi _{2,\mathcal{R}}\rangle $ could
be observed without difficulties in current experiments. This is illustrated
in figure \ref{fig1}a where we use the parameters $\omega _{0}/2\pi =8\,$%
GHz, $g_{0}=5\times 10^{-2}\omega _{0}$, $\Delta _{-}=0$, $\varepsilon
_{\Omega }=5\times 10^{-2}\Omega _{0}$, $\eta ^{(r)}=2\omega _{0}+g_{0}\sqrt{%
2}$ and consider moderate dissipation rates $\kappa =\gamma =\gamma _{\phi
}=2\times 10^{-4}g_{0}$. We plot the average photon number $\langle \hat{n}%
\rangle $, the Mandel factor $Q=[\langle (\Delta \hat{n})^{2}\rangle
-\langle \hat{n}\rangle ]/\langle \hat{n}\rangle $ (that quantifies the
spread of the photon number distribution) and the atomic excitation
probability $P_{e}$. We see that both models of dissipation predict
identical results and several oscillations of $\langle \hat{n}\rangle $, $Q$
and $P_{e}$ with high visibility could be observed within the time interval
of $500$\thinspace ns.

\subsection{Anti-Jaynes-Cummings (AJC) regime}

\label{s4a}

Now we consider the photon generation from ZES in the dispersive regime, $%
|\Delta _{-}|/2\gg g_{0}\sqrt{n_{\max }}$, where $n_{\max }$ is the maximum
number of excitations. The resonant modulation frequency is
\begin{equation}
\eta ^{(J)}=\Delta _{+}-2\left( \delta _{-}-\delta _{+}\right)
\end{equation}%
where $\delta _{-}$ ($\delta _{+}$) is the standard dispersive
(Bloch-Siegert) shift given by $\delta _{\pm }=g_{0}^{2}/\Delta _{\pm }$ and
$\Delta _{+}\equiv (\omega _{0}+\Omega _{0})$ (we neglected higher order
corrections to $\eta ^{(J)}$ \cite{I2}). The unitary dynamics is described
by the Anti-Jaynes-Cummings (AJC) Hamiltonian \cite{arxiv,jpcs,roberto},
which in the dressed picture reads $\tilde{H}=\theta |\varphi _{0,\mathcal{D}%
}\rangle \langle \varphi _{2,\mathcal{-D}}|+h.c.$,%
\begin{equation}
\theta =i\frac{1}{2}g_{0}\mathcal{D}\Upsilon ^{(J)}  \label{chick}
\end{equation}%
\begin{equation}
\Upsilon ^{(J)}=-\frac{\varepsilon _{\omega }^{(J)}}{\Delta _{+}}-\frac{%
\varepsilon _{\Omega }^{(J)}}{\Delta _{+}}+\frac{\varepsilon _{g}^{(J)}}{%
g_{0}}+i\frac{2\varepsilon _{\chi }^{(J)}}{\Delta _{-}}~.  \label{hob2}
\end{equation}%
Here $\mathcal{D}\equiv \Delta _{-}/\left\vert \Delta _{-}\right\vert =\pm $
is the \textquotedblleft detuning symbol\textquotedblright\ and we denote $%
|\varphi _{0,\mathcal{D}}\rangle \equiv |\varphi _{0,-}\rangle $.

For SME we find that under pure atomic dephasing the asymptotic solution is $%
\rho _{0}^{-}=\rho _{2}^{\pm }=\frac{1}{3}$, with all other probabilities
equal to zero. In the presence of pure atomic and cavity dampings we get the
asymptotic nonzero probabilities (to the second order in $g_{0}/\Delta _{-}$)%
\begin{equation}
\rho _{2}^{-\mathcal{D}}=\frac{\kappa \left( 1-\frac{g_{0}^{2}}{\Delta
_{-}^{2}}\right) +\gamma \frac{g_{0}^{2}}{\Delta _{-}^{2}}}{\kappa \frac{%
g_{0}^{2}}{\Delta _{-}^{2}}+\gamma \left( 1-3\frac{g_{0}^{2}}{\Delta _{-}^{2}%
}\right) }\rho _{1}^{\mathcal{D}}
\end{equation}%
\begin{equation}
\rho _{1}^{-\mathcal{D}}=\frac{\kappa \left( 1+\frac{g_{0}^{2}}{\Delta
_{-}^{2}}\right) +\gamma \frac{g_{0}^{2}}{\Delta _{-}^{2}}}{\kappa \frac{%
g_{0}^{2}}{\Delta _{-}^{2}}+\gamma \left( 1-\frac{g_{0}^{2}}{\Delta _{-}^{2}}%
\right) }\rho _{2}^{-\mathcal{D}}
\end{equation}%
\begin{equation}
\rho _{0}^{-}=\left[ 1+\left( \frac{\kappa +\gamma +2\frac{g_{0}^{2}}{\Delta
_{-}^{2}}\left( \kappa -\gamma \right) }{2\left\vert \theta \right\vert }%
\right) ^{2}\right] \rho _{2}^{-\mathcal{D}},
\end{equation}%
where $\rho _{0}^{-}$ can be easily found from the normalization condition.
Once again the dissipation changes drastically the dynamics, but for $\kappa
,\gamma \ll |\theta |$ excitations are still generated. In particular, for
pure atomic (cavity) damping and $\gamma \ll |\theta |$ ($\kappa \ll |\theta
|$) we obtain asymptotically $\rho _{1}^{\mathcal{D}}\approx 1$ ($\rho
_{1}^{-\mathcal{D}}\approx 1$), so only one photon (one atomic excitation)
is generated.

\begin{figure}[tbh]
\begin{center}
\includegraphics[width=.49\textwidth]{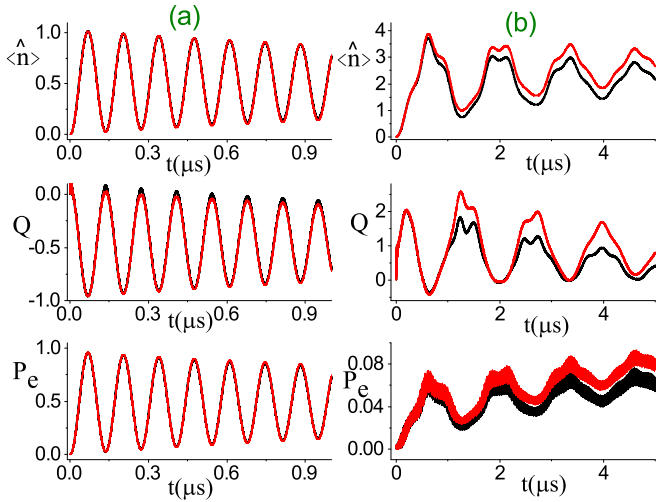} {}
\end{center}
\caption{Photon generation from ZES in the dispersive regime.
\textbf{(a)} AJC behavior (section \protect\ref{s4a}). \textbf{(b)} DCE
behavior (section \protect\ref{s3}). Black (red) lines stand for SME (DPME).}
\label{fig2}
\end{figure}

For our tentative parameters we obtain $|\theta |\sim 10^{-3}g_{0}$, of the
same order of magnitude as in section \ref{s4c}, so this effect could also
be implemented in circuit QED with current technology. This is illustrated
in figure \ref{fig2}a for parameters $\omega _{0}/2\pi =8\,$GHz, $%
g_{0}=5\times 10^{-2}\omega _{0}$, $\Delta _{-}=8g_{0}$, $\varepsilon
_{\Omega }=5\times 10^{-2}\Omega _{0}$, $\eta ^{(J)}=\lambda _{2,-}-\lambda
_{0}+2\delta _{+}\times 0.954$ and $\kappa =\gamma =\gamma _{\phi
}=g_{0}\times 10^{-4}$. Once again the results are almost indistinguishable
for the two dissipation models and the periodic generation of one cavity and
one atomic excitations is easily observable on timescales of the order of $%
500\,$ns. We notice that two excitations can also we generated from
nonvacuum initial states for $\eta ^{(J)}=\Delta _{+}-2k(\delta _{-}-\delta
_{+})$, where $k$ is an integer and the corresponding generation rate in
equation (\ref{chick}) becomes $\theta \sqrt{k}$ \cite{jpcs,I2}.

\subsection{Anti-DCE regime}

\label{s4b}

The Anti-DCE behavior was predicted recently in \cite{I2}. It occurs in the
dispersive regime and consists of coherent annihilation of two system
excitations via the coupling $|\varphi _{k,\mathcal{D}}\rangle
\leftrightarrow |\varphi _{k-2,-\mathcal{D}}\rangle $ for a given value of $%
k\geq 3$, roughly equivalent to the transition $|g,k\rangle \leftrightarrow
|e,k-3\rangle $. For the initial state $|g\rangle \langle g|\otimes \hat{\rho%
}_{\mathrm{field}}$ three photons are subtracted from the field due to the
time-modulation of the system parameters, so the denomination Anti-DCE seems
appropriate. Anti-DCE is implemented for the modulation frequency%
\begin{equation}
\eta _{k}^{(A)}=3\omega _{0}-\Omega _{0}+2\allowbreak \left( \delta
_{-}-\delta _{+}\right) \left( k-1\right) \,,  \label{etaka}
\end{equation}%
where $k\geq 3$ and the dressed-picture Hamiltonian is $\tilde{H}=\theta
^{\ast }|\varphi _{k,\mathcal{D}}\rangle \langle \varphi _{k-2,-\mathcal{D}%
}|+h.c.$,%
\begin{equation}
\theta =i\mathcal{D}\frac{\delta _{-}\Omega _{0}g_{0}}{2\omega _{0}\Delta
_{-}}\sqrt{k\left( k-1\right) (k-2)}\Upsilon ^{(A)}  \label{tetaADCE}
\end{equation}%
\begin{equation}
\Upsilon ^{(A)}=\frac{\varepsilon _{\omega }^{(A)}}{2\omega _{0}+\Delta _{-}}%
+\frac{\omega _{0}+\Delta _{-}}{2\omega _{0}+\Delta _{-}}\frac{\varepsilon
_{\Omega }^{(A)}}{\Omega _{0}}-\frac{\varepsilon _{g}^{(A)}}{g_{0}}\,.
\label{hob3}
\end{equation}%
This phenomenon occurs only for the modulation of $\omega $, $\Omega $ or $g$%
, so it cannot be implemented by the parametric down-conversion process in
which only $\varepsilon _{\chi }\neq 0$. For our tentative parameters, $%
|\Delta _{-}|\sim 10g_{0}$ and $k\sim 5$, we get $|\theta |\sim 10^{-4}g_{0}$%
, so Anti-DCE is a rather weak effect that requires fine-tuning of the
modulation frequency and a prolonged maintenance of perturbation. Still, the
transition rate $|\theta |$ is slightly greater than the state-of-the-art
dissipation rates, so this effect lies on the threshold of implementability
with current technology.

For SME, under pure atomic dephasing we obtain asymptotically $\rho _{k}^{%
\mathcal{\pm }}=\rho _{k-2}^{\mathcal{\pm }}$ and $\rho _{l}^{\mathcal{+}%
}=\rho _{l}^{-}$ for $l>0$, so for large times the dephasing changes
completely the expected behavior. Anti-DCE behavior is also strongly
affected by the cavity relaxation, when one gets approximately%
\begin{eqnarray}
\frac{d\tilde{\rho}_{N}^{\mathcal{D}}}{d(\kappa t)} &=&\left( N+1\right)
\left( 1-\frac{g_{0}^{2}}{\Delta _{-}^{2}}\right) \tilde{\rho}_{N+1}^{%
\mathcal{D}} \\
&&+\frac{g_{0}^{2}}{\Delta _{-}^{2}}\tilde{\rho}_{N+1}^{-\mathcal{D}%
}-N\left( 1-\frac{g_{0}^{2}}{\Delta _{-}^{2}}\right) \tilde{\rho}_{N}^{%
\mathcal{D}}  \nonumber
\end{eqnarray}%
\begin{eqnarray}
\frac{d\tilde{\rho}_{N,N}^{\mathcal{-D}}}{d(\kappa t)} &=&N\left( 1+\frac{%
g_{0}^{2}}{\Delta _{-}^{2}}\right) \tilde{\rho}_{N+1}^{-\mathcal{D}} \\
&&-\left( N-1+\frac{g_{0}^{2}}{\Delta _{-}^{2}}N\right) \tilde{\rho}_{N}^{-%
\mathcal{D}}.  \nonumber
\end{eqnarray}%
Similar formulae hold for the pure atomic damping:%
\begin{eqnarray}
\frac{d\tilde{\rho}_{N}^{\mathcal{D}}}{d(\gamma t)} &=&\left( 1-\frac{%
g_{0}^{2}}{\Delta _{-}^{2}}\left( 2N+1\right) \right) \tilde{\rho}_{N+1}^{-%
\mathcal{D}} \\
&&+\frac{g_{0}^{2}}{\Delta _{-}^{2}}\left( N+1\right) \tilde{\rho}_{N+1}^{%
\mathcal{D}}-\frac{g_{0}^{2}}{\Delta _{-}^{2}}N\tilde{\rho}_{N}^{\mathcal{D}}
\nonumber
\end{eqnarray}%
\begin{equation}
\frac{d\tilde{\rho}_{N}^{-\mathcal{D}}}{d(\gamma t)}=\frac{g_{0}^{2}}{\Delta
_{-}^{2}}N\tilde{\rho}_{N+1}^{-\mathcal{D}}-\left( 1-\frac{g_{0}^{2}}{\Delta
_{-}^{2}}N\right) \tilde{\rho}_{N}^{-\mathcal{D}}.
\end{equation}%
In both cases the system state goes asymptotically to ZES. This did not
happen in the phenomena analyzed in sections \ref{s4c} and \ref{s4a} because
there the ZES was coupled to $2$-excitations states by the modulation,
resulting in nonvacuum equilibrium state.
\begin{figure}[tbh]
\begin{center}
\includegraphics[width=.49\textwidth]{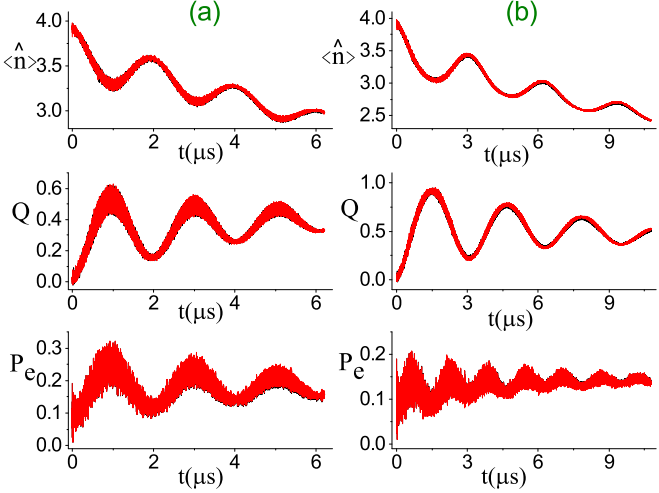} {}
\end{center}
\caption{Coherent photon annihilation in the dispersive regime for initial
coherent state $|g\rangle \otimes |\protect\alpha \rangle $. \textbf{(a)}
Anti-DCE behavior (section \protect\ref{s4b}). \textbf{(b)} Enhanced
Anti-DCE behavior (section \protect\ref{s5b}). Black (red) lines stand for
SME (DPME).}
\label{fig3}
\end{figure}

Figure \ref{fig3}a assesses the feasibility of experimental observation of
Anti-DCE for the initial coherent state $|g\rangle \otimes |\alpha \rangle $%
. We use the values $\alpha =2$, $\omega _{0}/2\pi =8\,$GHz, $g_{0}=5\times
10^{-2}\omega _{0}$, $\Delta _{-}=8g_{0}$, $\varepsilon _{\Omega }=5\times
10^{-2}\Omega _{0}$, $\kappa =10^{-5}g_{0}$, $\gamma =\gamma _{\phi
}=5\times 10^{-5}g_{0}$ and $\eta _{4}^{(A)}=\lambda _{4,+}-\lambda
_{2,-}-2\delta _{+}\times 2.791$. Both master equations predict practically
identical results and for initial times the system exhibits the expected
behavior, characterized by the periodic reduction of the average photon
number accompanied by the partial excitation of the atom and modification of
the field statistics from Poissonian to super-Poissonian. The curves
corresponding to $\langle \hat{n}\rangle $, $Q$ and $P_{e}$ exhibit broad
widths due to fast oscillations associated with off-resonant exchange of
excitations between the atom and the field that occurs on timescales ($\sim
|\Delta _{-}|^{-1}$) much smaller that the characteristic time of Anti-DCE.
For larger times the system tends to the ZES and the amplitude of
oscillations, as well as the associated visibility, decrease. Yet the main
features of Anti-DCE could be resolved in experiments with the current
state-of-the-art parameters on timescales of a few $\mu $s.

\section{Four-excitations behavior}

\label{s5}

\subsection{Enhanced photon generation from vacuum in resonant regime}

\label{s5a}

Now we consider the two-tone multi-modulation applied simultaneously to $%
\Omega $ and $g$, with one frequency given by equation (\ref{fre1}) and the
second%
\begin{equation}
\eta ^{(r2)}=2\omega _{0}+g_{0}\sqrt{2}(\sqrt{2}\mathcal{R}_{2}-\mathcal{R)}%
~,~\mathcal{R},\mathcal{R}_{2}=\pm ~.
\end{equation}%
For initial ZES the effective Hamiltonian reads $\tilde{H}=\theta |\varphi
_{0,-}\rangle \langle \varphi _{2,\mathcal{R}}|+\theta _{2}|\varphi _{2,%
\mathcal{R}}\rangle \langle \varphi _{4,\mathcal{R}_{2}}|+h.c.$, where $%
\theta $ is given by (\ref{teta1}) and%
\begin{equation}
\theta _{2}=ig_{0}\mathcal{R}_{2}\frac{\sqrt{3}}{4}\Upsilon ^{(r2)}~
\end{equation}%
\begin{equation}
\Upsilon ^{(r2)}=\frac{\varepsilon _{\omega }^{(r2)}}{2\omega _{0}}+\frac{%
\varepsilon _{\Omega }^{(r2)}}{2\omega _{0}}-\frac{\varepsilon _{g}^{(r2)}}{%
g_{0}}+i\mathcal{R}_{2}\frac{\varepsilon _{\chi }^{(r2)}}{g_{0}}(2+\mathcal{%
RR}_{2}\sqrt{2})~.
\end{equation}

It is worth writing first the solution for the unitary evolution. We expand
the interaction-picture wavefunction as $|\tilde{\psi}\rangle =A|\varphi
_{0,-}\rangle +B|\varphi _{2,\mathcal{R}}\rangle +C|\varphi _{4,\mathcal{R}%
_{2}}\rangle +\sum (\cdots )$, where $\sum (\cdots )$ denotes the other
terms not coupled by the Hamiltonian and we assume real probability
amplitudes. For the initial condition $B_{0}=0$ we find%
\begin{eqnarray}
A &=&\left( A_{0}+\frac{\theta _{2}}{\theta ^{\ast }}C_{0}\right) \cos Rt+%
\frac{\theta _{2}}{R^{2}\theta ^{\ast }}\left[ 2\theta ^{\ast }\theta
_{2}^{\ast }A_{0}\right.  \label{f1} \\
&&\left. -2\theta _{2}^{\ast }(\theta ^{\ast }A_{0}+\theta _{2}C_{0})\cos
^{2}\frac{1}{2}Rt-(|\theta |^{2}-|\theta _{2}|^{2})C_{0}\right]  \nonumber
\end{eqnarray}%
\begin{equation}
B=-i\frac{(\theta ^{\ast }A_{0}+\theta _{2}C_{0})\sin Rt}{R}
\end{equation}%
\begin{eqnarray}
C&=&\frac{2\theta _{2}^{\ast }(\theta ^{\ast }A_{0}+\theta _{2}C_{0})\cos ^{2}%
\frac{1}{2}Rt}{R^{2}} \nonumber \\
&&+\frac{(|\theta |^{2}-|\theta _{2}|^{2})C_{0}-2\theta
^{\ast }\theta _{2}^{\ast }A_{0}}{R^{2}},  \label{f3}
\end{eqnarray}%
where $R\equiv \sqrt{|\theta |^{2}+|\theta _{2}|^{2}}$. One can transfer the
populations entirely from $\{A$,$C\}$ to $C$ at the smallest time $Rt_{\min
}=\pi $ by choosing $\theta _{2}=x\theta ^{\ast }$, $x=\left( C_{0}\pm \sqrt{%
C_{0}^{2}+A_{0}^{2}}\right) /A_{0}$. In the present case $x=\pm 1$, so one
can create the pure dressed state $|\varphi _{4,\mathcal{R}_{2}}\rangle $ at
the time $t_{\min }=\pi /(|\theta |\sqrt{2})$.

For the pure atomic dephasing described by SME we get in the asymptotic
limit $\rho _{0}^{-}=\rho _{2}^{\pm }=\rho _{4}^{\pm }=1/5$, while for the
pure atomic damping the nonzero probabilities are%
\begin{equation}
\rho _{0}^{-}=\left( \frac{2\left\vert \theta \right\vert ^{2}+\left( \gamma
/4\right) ^{2}}{\left\vert \theta \right\vert ^{2}}-2\frac{\left\vert \theta
\right\vert ^{2}-\left( \gamma /4\right) ^{2}}{3\left\vert \theta
\right\vert ^{2}+\gamma ^{2}/4}\right) \rho _{2}^{\mathcal{R}}
\end{equation}%
\begin{equation}
\rho _{4}^{\mathcal{R}_{2}}=\frac{4\left\vert \theta \right\vert ^{2}}{%
3\left\vert \theta \right\vert ^{2}+\gamma ^{2}/4}\rho _{2}^{\mathcal{R}}
\end{equation}%
\begin{equation}
\rho _{1}^{\pm }=\frac{1}{2}\frac{5\left\vert \theta \right\vert ^{2}+\gamma
^{2}/4}{3\left\vert \theta \right\vert ^{2}+\gamma ^{2}/4}\rho _{2}^{%
\mathcal{R}}
\end{equation}%
\begin{equation}
\rho _{3}^{\pm }=\rho _{2}^{\mathcal{-R}}=\frac{1}{2}\rho _{4}^{\mathcal{R}%
_{2}}~,
\end{equation}%
where $\rho _{0}^{-}$ can be found from the normalization condition. Similar
expressions are obtained for the pure cavity damping, so we conclude that
for $\kappa ,\gamma \ll |\theta |$ four excitations can be generated from
vacuum. For our tentative parameters we get $|\theta |,|\theta _{2}|\sim
10^{-3}g_{0}$, so this effect can also be implemented with current
technology.

We illustrate the typical outcome in figure \ref{fig1}b, where we consider
the modulation of $\Omega $ with frequency $\eta ^{(r)}=2\omega _{0}+g_{0}%
\sqrt{2}$ and modulation of $g$ with frequency $\eta ^{(r2)}=2\omega
_{0}+g_{0}(2-\sqrt{2})$. Other parameters are $\omega _{0}/2\pi =8\,$GHz, $%
g_{0}=5\times 10^{-2}\omega _{0}$, $\Delta _{-}=0$, $\varepsilon _{\Omega
}=5\times 10^{-2}\Omega _{0}$, $\varepsilon _{g}=1.97\times 10^{-2}g_{0}$, $%
w_{\Omega }^{(r)}=w_{g}^{(r2)}=1$ and $\kappa =\gamma =\gamma _{\phi
}=2\times 10^{-4}g_{0}$. We see that even for moderate values of dissipation
rates the enhanced photon generation due to multi-modulation can still be
observed. Moreover, the predictions of SME and DPME are almost
indistinguishable.

\subsection{Photon annihilation in enhanced Anti-DCE regime}

\label{s5b}

One can enhance the Anti-DCE behavior by combining the modulation frequency (%
\ref{etaka}) with the second frequency
\begin{equation}
\eta _{k}^{(A2)}=\Delta _{+}-2\left( \delta _{-}-\delta _{+}\right) \left(
k-3\right) \,,  \label{AJC}
\end{equation}%
where $k\geq 4$ is an integer. The effective Hamiltonian becomes $\tilde{H}%
=\theta ^{\ast }|\varphi _{k,\mathcal{D}}\rangle \langle \varphi _{k-2,-%
\mathcal{D}}|+\theta _{2}^{\ast }|\varphi _{k-2,-\mathcal{D}}\rangle \langle
\varphi _{k-4,\mathcal{D}}|+h.c.$, where $\theta $ is given by (\ref%
{tetaADCE}) and
\begin{equation}
\theta _{2}=i\frac{1}{2}g_{0}\mathcal{D}\sqrt{k-3}\Upsilon ^{(A2)}
\end{equation}%
\begin{equation}
\Upsilon ^{(A2)}=-\frac{\varepsilon _{\omega }^{(A2)}}{\Delta _{+}}-\frac{%
\varepsilon _{\Omega }^{(A2)}}{\Delta _{+}}+\frac{\varepsilon _{g}^{(A2)}}{%
g_{0}}+i\frac{2\varepsilon _{\chi }^{(A2)}}{\Delta _{-}}\,.
\end{equation}

In this regime one combines simultaneously the coupling $|\varphi _{k,%
\mathcal{D}}\rangle \longrightarrow |\varphi _{k-2,-\mathcal{D}}\rangle $
via Anti-DCE with the coupling $|\varphi _{k-2,-\mathcal{D}}\rangle
\rightarrow |\varphi _{k-4,\mathcal{D}}\rangle $ via the AJC behavior,
obtaining the annihilation of four excitations. In the dispersive regime
this corresponds to subtraction of four photons from the state $|g,k\rangle
\langle g,k|$. The effective Hamiltonian is similar to the one considered in
section \ref{s5a}, so the formulae (\ref{f1}) -- (\ref{f3}) can be applied
after appropriate substitutions. If the initial cavity state is known and
the population of the excited atomic state is negligible, one can transfer
completely the populations of $\{|\varphi _{k,\mathcal{D}}\rangle ,|\varphi
_{k-4,\mathcal{D}}\rangle \}$ to $|\varphi _{k-4,\mathcal{D}}\rangle $ at
specific times without affecting the other states, so four photons are
indeed annihilated. If the initial statistics is not known, the transfer is
not complete because the required value of $\theta _{2}$ cannot be
determined beforehand.

For SME we reach the same conclusions about the asymptotic behavior as in
section \ref{s4b}, supplemented by the additional condition $\rho _{k}^{%
\mathcal{D}}=\rho _{k-2}^{\mathcal{D}}=\rho _{k-4}^{\mathcal{D}}$ under pure
atomic dephasing. As the associated transition rates are similar to the one
estimated in section \ref{s4b}, the enhanced Anti-DCE also lies at the
threshold of experimental feasibility, provided two simultaneous fine-tuned
modulations can be sustained for a time interval of a few $\mu $s. This is
illustrated in figure \ref{fig3}b for the initial coherent state $|g\rangle
\otimes |\alpha \rangle $ and the simultaneous modulation of $\Omega $ and $g
$. We used the parameters of figure \ref{fig3}a, $\varepsilon
_{g}=9.91\times 10^{-4}g_{0}$, $\eta _{4}^{(A2)}=\lambda _{2,-}-\lambda
_{0}+2\delta _{+}\times 0.954$ and nonzero modulation weights $w_{\Omega
}^{(A)}=w_{g}^{(A2)}=1$. We see that the reduction of $\langle \hat{n}%
\rangle $ is stronger than in the standard Anti-DCE case, occurring at
larger timescales; the $Q-$factor increases while the atomic excitation
probability undergoes faster oscillations and is slightly smaller than in
figure \ref{fig3}a because the transfer of population from $|g,4\rangle $ to
$|e,1\rangle $ is only partial. Moreover, SME and DPME\ predict almost
identical behaviors, with minor quantitative differences appearing only for
larger times $t\gtrsim 5\,\mu $s.

\section{Dynamical Casimir effect (DCE) with single qubit}

\label{s3}

In the dispersive regime one can create many pairs of excitations from the
initial ZES for the modulation frequency%
\begin{equation}
\eta ^{(d)}=2(\omega _{0}+\allowbreak \delta _{-}-\delta _{+})\,.
\end{equation}%
The dynamics is described by the effective Hamiltonian
\begin{equation}
\tilde{H}=\sum_{m=0}^{\infty }\theta _{m}e^{-it(\lambda _{m+2,\mathcal{D}%
}-\lambda _{m,\mathcal{D}}-\eta ^{(d)})}|\varphi _{m,\mathcal{D}}\rangle
\langle \varphi _{m+2,\mathcal{D}}|+h.c.,  \label{sat}
\end{equation}%
\begin{equation}
\theta _{m}=i\frac{\delta _{-}\Omega _{0}}{2\Delta _{+}}\sqrt{(m+1)(m+2)}%
\Upsilon ^{(d)}
\end{equation}%
\begin{equation}
\Upsilon ^{(d)}=\frac{\varepsilon _{\omega }^{(d)}}{\omega _{0}}+\frac{%
\varepsilon _{\Omega }^{(d)}}{\Omega _{0}}-2\frac{\varepsilon _{g}^{(d)}}{%
g_{0}}+i\frac{\Delta _{+}}{\Omega _{0}}\frac{\varepsilon _{\chi }^{(d)}}{%
\delta _{-}}\,,  \label{hob4}
\end{equation}%
so only the states $|\varphi _{2m,\mathcal{D}}\rangle \simeq |g,2m\rangle $
are populated during the dissipationless evolution. The unitary dynamics of
the Hamiltonian was studied in \cite{I2}, where it was shown that the photon
generation from vacuum suffers saturation due to effective Kerr nonlinearity
(of magnitude $g_{0}^{4}/\Delta _{-}^{3}$), and $\langle \hat{n}\rangle $
exhibits a collapse--revival behavior as function of time. Simple analytical
expressions cannot be obtained for this case because many dressed states are
coupled simultaneously and the argument of the exponentials depends on the
index $m$ in nontrivial manner.

For SME the pure atomic damping is equivalent to the pure cavity damping
with effective relaxation rate $\kappa _{ef}=\gamma (g_{0}/\Delta _{-})^{2}$
(for the initial ZES and under RWA). The asymptotic solution for a similar
problem (parametric amplification in the presence of Kerr nonlinearity and
cavity relaxation) was obtained two decades ago using the method of
potential solutions for the corresponding Fokker-Planck equation \cite%
{kryuch1,kryuch2}. However, the asymptotic solution is of little use in our
case because the photon statistics during the transient time can be quite
different from the asymptotic one \cite{I2}, so we rely entirely on
numerical simulations.

For our tentative parameters we estimate $|\theta _{0}^{(d)}|\sim
10^{-4}g_{0}$, so in principle the single-qubit DCE could be observed in the
current state-of-the-art architectures. This is confirmed in figure \ref%
{fig2}b where we show the expected behavior for parameters $\omega _{0}/2\pi
=8\,$GHz, $g_{0}=5\times 10^{-2}\omega _{0}$, $\Delta _{-}=8g_{0}$, $%
\varepsilon _{\Omega }=5\times 10^{-2}\Omega _{0}$, $\kappa =\gamma =\gamma
_{\phi }=5\times 10^{-5}g_{0}$ and $\eta ^{(d)}=\lambda _{2,+}-\lambda
_{0}-2\delta _{+}\times 1.02$. Both dissipation models predict identical
results for initial times; for times $t\gtrsim 1\,\mu $s some small
quantitative differences appear without changing the overall behavior. The
saturation in photon growth occurs due to the nonzero values of $(\lambda
_{m+2,\mathcal{D}}-\lambda _{m,\mathcal{D}}-\eta ^{(d)})\propto m$ in (\ref%
{sat}) for large values of $m$, and is not related to the presence
dissipation. We see that photons can be generated and the collapse--revival
of $\langle \hat{n}\rangle $ can be observed on the timescale of a few $\mu $%
s. Notice that the collapse--revival behavior is not associated to the
absorption and reemission of photons by the atom \cite{milburn1,milburn2},
as the atomic excitation probability stays $\lesssim 8\%$ for all times.
More photons can be generated if one adds a second modulation frequency or
increases $|\Upsilon ^{(d)}|$ or $\Delta _{-}$ \cite{I2}, but in the latter
case the parameter $|\theta _{m}|$ decreases so the process becomes slower
and demands more precise tuning of the modulation frequency.

\section{Conclusions}

\label{s6}

We estimated the rates of photon creation and annihilation for some
nonstationary phenomena induced by fast modulation of qubit parameters in
dissipative circuit QED. Two different Markovian master equations
(\textquotedblleft standard\textquotedblright\ and \textquotedblleft
dressed-picture\textquotedblright\ master equations) were used to account
for the common sources of dissipation, and we verified that for the relevant
regime of parameters the predicted behaviors are almost identical.

Our results indicate that all the analyzed phenomena could be implemented
experimentally with the present technology, provided one can implement
modulations with relative depths $|\Upsilon |\sim 10^{-2}$ and frequencies $%
\eta \sim 15\,$GHz. The most accessible phenomena (\textquotedblleft fast
phenomena\textquotedblright ) are the two- and four-excitations generation
from vacuum in the resonant regime (using single- and two-tone modulations,
respectively) and the two-excitation generation via the AJC behavior in the
dispersive regime. For these effects the photon generation rates are $%
|\theta |\sim 10^{-3}g_{0}$, so they could be implemented even in systems
with moderate dissipation. On the other hand, generation of several photons
from vacuum via DCE and annihilation of two or four excitations via Anti-DCE
in the dispersive regime (\textquotedblleft slow phenomena\textquotedblright
) occur at lower rates $|\theta |\sim 10^{-4}g_{0}$, so only
state-of-the-art architectures would allow for their experimental
verification. Additionally, the modulation frequency must be fine-tuned with
accuracy of the order of $|\theta |$ and maintained for the time interval $%
\sim 500\,$ns for the fast phenomena and $\sim 5\,\mu $s for the slow ones.
Although all these requirements are challenging, the verification of the
nonstationary phenomena in circuit QED could lead to new manners of
generating entangled states by means of the counter-rotating terms in the
light--matter interaction Hamiltonian.

\begin{acknowledgments}
DSV acknowledges financial support by CAPES (Brazilian agency). AVD
acknowledges partial support by CNPq, Conselho Nacional de Desenvolvimento
Cient\'{\i}fico e Tecnol\'{o}gico -- Brazil.
\end{acknowledgments}

\end{document}